\documentclass[5p]{elsarticle}
\usepackage{epstopdf}
\usepackage{lineno}
\usepackage{color}
\usepackage{listings}
\usepackage{dirtree}
\usepackage{hyperref}

\modulolinenumbers[5]

\newcommand{\linka}[1]{\href{#1}{#1}}

\newcommand{\x}{}
\journal{Journal of Astronomy and Computing}

\biboptions{authoryear}

\begin{document}
\begin{frontmatter}
\title{A web portal for hydrodynamical, cosmological simulations\tnoteref{mytitlenote}}
\tnotetext[mytitlenote]{access via \href{https://c2papcosmosim.uc.lrz.de}{https://c2papcosmosim.uc.lrz.de}.}

\author[excluster,lrz,usm]{Ragagnin, A. \fnref{myfootnote}}
\fntext[myfootnote]{ragagnin@lrz.de}

\author[usm,mpa]{Dolag, K.}
\author[auts,inaf]{Biffi, V.}
\author[excluster,rzg]{Cadolle Bel, M.}
\author[lrz]{Hammer, N.~J.}
\author[lrz,excluster]{Krukau, A.}
\author[usm,excluster]{Petkova, M.}
\author[lrz]{\\ and Steinborn, D.}

\address[lrz]{Leibniz-Rechenzentrum (LRZ), Boltzmannstrasse 1, D-85748 Garching, Germany}
\address[excluster]{Excellence Cluster Universe, Boltzmannstrasse 2r, D-85748 Garching, Gemany}
\address[usm]{Universit\"ats-Sternwarte, Fakult\"at f\"ur Physik, Ludwig-Maximilians Universit\"at M\"unchen, Scheinerstrasse 1, D-81679 M\"unchen, Germany}
\address[mpa]{Max-Planck-Institut f\"ur Astrophysik, Karl-Schwarzschild Strasse 1,D-85748 Garching bei M\"unchen, Germany}
\address[auts]{Astronomy Unit, Department of Physics, University of Trieste, via Tiepolo 11, I-34131 Trieste, Italy}
\address[inaf]{INAF, Osservatorio Astronomico di Trieste, via Tiepolo 11, I-34131 Trieste, Italy}
\address[rzg]{Max Planck Computing and Data Facility (MPCDF), Gie$\beta$enbachstrasse 2, D-85748 Garching, Germany}

\begin{abstract}
{This article describes a data center {hosting a web portal} for accessing and sharing
the output of large, cosmological, hydro-dynamical simulations
with a broad scientific community. It also allows users to
receive related scientific data products by directly processing the
raw simulation data on a remote computing cluster.}

{The data center has a multi-layer structure: a web portal,
a job control layer, a computing cluster and a HPC
storage system. The outer layer enables users to choose an
object from the simulations. Objects can be selected by visually
inspecting 2D maps of the simulation data, by performing highly compounded and elaborated queries or graphically by plotting
arbitrary combinations of properties.
 {The user can run analysis tools on a chosen object.}
These services allow users to run analysis tools on the raw simulation data.
The job control layer is responsible for handling and performing the analysis jobs, which are executed
on a computing cluster. The innermost layer
is formed by a HPC storage system which hosts the large, raw simulation data.}

{The following services are available for the users:
(I) {\sc ClusterInspect} visualizes properties of member
galaxies of a selected galaxy cluster; (II) {\sc SimCut} returns
the raw data of a sub-volume around a selected object from a
simulation, containing all the original, hydro-dynamical quantities;
(III) {\sc Smac} creates idealised 2D maps of various, physical
quantities and observables of a selected object; (IV) {\sc Phox} generates
virtual X-ray observations with specifications of various current and
upcoming instruments.}
\end{abstract}


\end{frontmatter}


\section{Introduction}

\begin{figure*}
\includegraphics[width=0.99\textwidth]{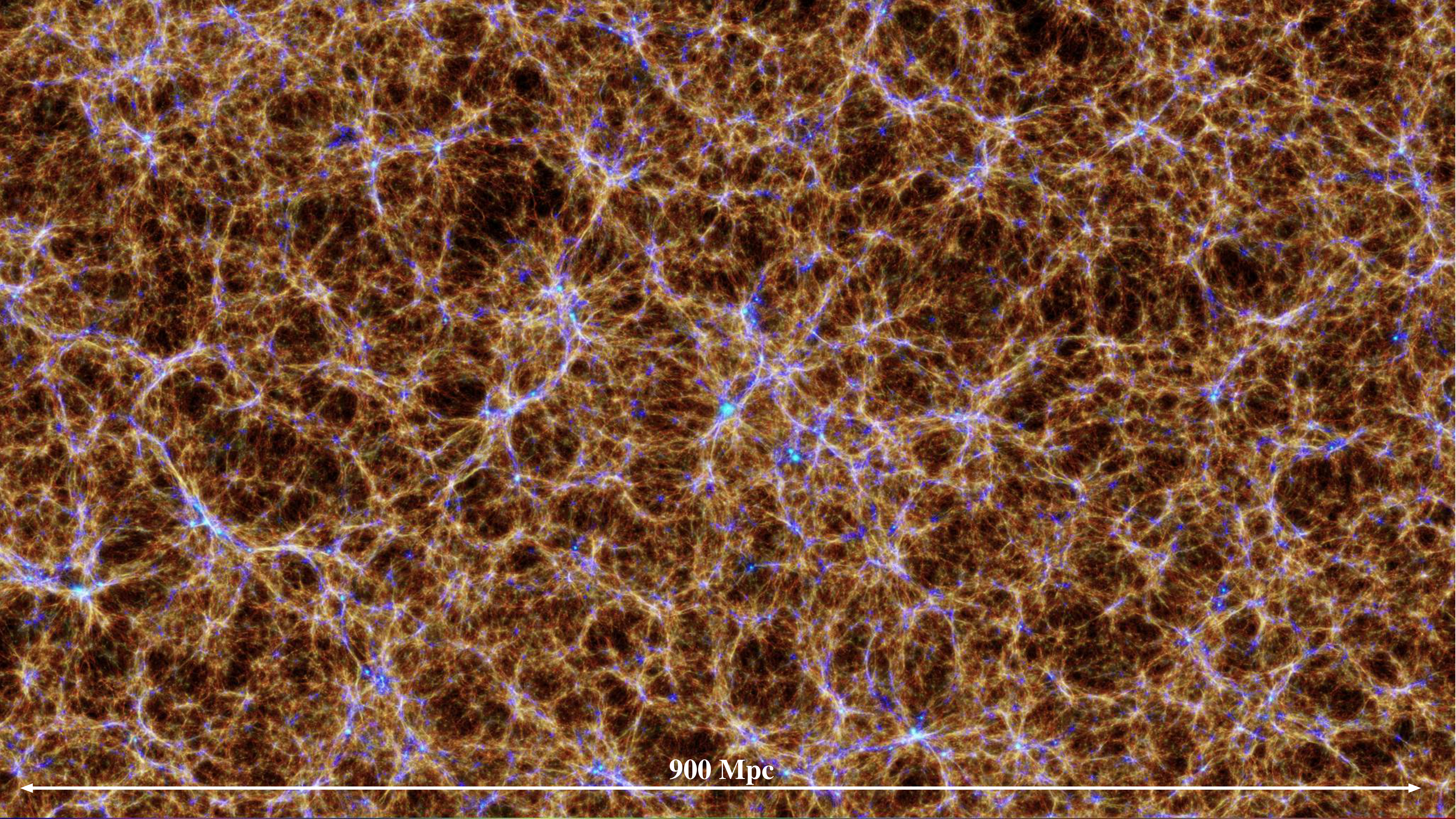}
\caption{\x{A visualisation of a cosmological large scale structure of the {\it Box2b/hr}  simulation from the {\it Magneticum} project. This map shows  diffuse baryons at $z=0.2,$ colour coded according to their temperature. The visualisation is centred on the most massive galaxy
cluster in this simulation.}}
\label{fig:box2b}
\end{figure*}

Entering the so-called era of ``precision cosmology'' it {becomes} more and
more clear that a theoretical counterpart {in the form of} very complex,
{hydrodynamical cosmological simulations is needed to interpret}
data from upcoming astronomical surveys and current instruments like PLANCK,
South Pole Telescope (SPT), PanStars, Dark Energy Survey (DES),
Euclid, LOFAR, eROSITA and many more. Such
simulations follow the growth of
galaxies and their associated components (like stellar
population and central black hole) with their interplay with
the large scale environment they are embedded in. Upcoming surveys
will map large volumes of the Universe as well as record the birth of
the first structures, especially galaxies and even progenitors of
massive galaxy clusters at high redshift. In fact, their large potential
of determining the nature of dark matter and dark energy
comes from being able to map the content and geometry of the Universe
over most time in cosmic history. For theoretical models this
means that simulations have to cover comparable large volumes,
especially to host the rarest, most massive galaxy clusters expected to be the
lighthouses of structure formation detectable at high redshift.
While the Universe makes its transition from dark matter dominated to
dark energy dominated (i.e. accelerated expansion), the objects
which form within it make their transition from young, dynamically
active and star formation-driven systems to more relaxed and
\x{equilibrated systems observed at low redshifts.}
Those simulations study the internal evolution of
clusters of galaxies with respect to the evolution of the
cosmological background. They will be essential to
interpret the outstanding discoveries expected from upcoming
surveys.

\x{However, running, storing and analysing such simulations is }
a challenging task, both from a technical as well as
from a collaborative point of view. Recent generations of HPC
facilities provided within initiatives like GAUSS\footnote{\linka{https://gauss-allianz.de/}} or PRACE\footnote{\linka{http://www.prace-ri.eu/}}
belong to the first generation of supercomputers which
\x{can perform cosmological, hydrodynamical}
simulations  covering both the required large volume and high
resolution requirements. Here, the largest simulation performed so far
belongs to the {\it Magneticum} project\footnote{\linka{http://www.magneticum.org}} and follows
$2\times4536^3$ resolution elements over the whole, cosmological evolution
\x{of the universe \citep{2016MNRAS.456.2361B}.
Such simulations model many more physical processes (star formation, cooling, winds, etc.) than the typical dark matter only counterparts
used currently in computational cosmology.}
{ These simulations provide 
a larger set of complex data  and can reach several hundreds of terabytes of raw data.
Such simulations are performed within large collaborative}
efforts and results have to be  shared {with}  a broader scientific community.
A guarantee for a deep scientific impact means that such data are made easily
accessible and processable within the individual collaborating groups.
\x{It implies that data are stored on the HPC facilities for long periods of
time, with the possibility to post-process the full data.
In addition, it is important to {make such data available} to a large
astrophysical community { and allow the scientists to apply analysis tools} 
via standard interfaces.}

{ In this respect, efforts have been done in the recent years in order to share data sets of various kinds with the community. For instance,}
\x{the Millennium Simulation Data Archive\footnote{\linka{http://wwwmpa.mpa-garching.mpg.de/Millennium/}} \citep{2006astro.ph..8019L} is a pioneering work in this field.
With the Millennium Simulation Data Archive, the user is able to compose  {\sc SQL} queries over
substructure and merger-tree data in order to extract
haloes and galaxies from the Millennium Simulation.

Users can also download the raw data files.
The Cosmosim.org project\footnote{\linka{https://www.cosmosim.org/}}
allows
users to compose additional queries  over the list of particles and various post processed quantities (grid cells of density field).
The Illustris Galaxies Observatory\footnote{\linka{http://www.illustris-project.org/galaxy\_obs/}}
provides an application programming interface (API)
where users can filter galaxies and download particle data
from the Illustris simulations.
The Australian Theoretical Virtual Observatory\footnote{\linka{https://tao.asvo.org.au/tao/about/}}\citep{2016ApJS..223....9B}
{is an online virtual laboratory where users can compose queries and run services on selected objects in the simulation, for instance producing mock observations or extracting light cones.}
}

Section \ref{sim}
\x{describes data of cosmological simulations  and section \ref{concept}
describes the currently available infrastructure.
In section \ref{web} we describe how { users can interact with the web interface and
thereby} compose science-driven queries to select objects.}
\x{Section \ref{services} describes the  services currently implemented in the system.}

\section{The Simulations}
\label{sim}

\x{In this section we present the simulations made accessible by our  data center.}

\subsection{The Magneticum Project}

\x{The Magneticum simulations\footnote{\linka{http://www.magneticum.org}}
(see \citet{2013MNRAS.428.1395B,
2014MNRAS.440.2610S, 2014MNRAS.442.2304H, 2015MNRAS.448.1504S,
2015arXiv150905134D, 2015MNRAS.451.4277D, 2015ApJ...812...29T,
2016MNRAS.458.1013S, 2016MNRAS.456.2361B, 2016arXiv160301619R})
follow  the evolution of up to $2\times10^{11}$ particles in a
series of cosmological boxes ranging in size from (50Mpc)$^3$ to
(4Gpc)$^3.$  } \x{A visualisation} of the second largest
cosmological simulation can be seen in Figure \ref{fig:box2b}. These
simulations were used to interpret Sunyaev-Zel'dovich  (SZ) data from
PLANCK \citep{2013A&A...550A.131P} and SPT \citep{2014ApJ...794...67M}
as well as to predict cluster properties in X-rays for future missions
such as Athena or Astro-H \citep{2013MNRAS.428.1395B}. The first mock
observations for the eROSITA cluster working group and the Athena+
white book were also produced based on these simulations. Other
scientific goals that were achieved with these simulations included
studying the { properties of the intra cluster medium (ICM) in}
galaxy clusters \citep{2015arXiv150905134D} as well as predicting the
multi wavelength properties of the Active Galactic Nuclei (AGN) \citep{2014MNRAS.442.2304H,
  2015MNRAS.448.1504S}.
The large dynamical range probed by the
combination of resolution levels and cosmological volumes also allowed
us to calibrate the cosmological mass function based on
hydro-dynamical simulations to a level required by future cosmological
probes \citep{2016MNRAS.456.2361B}.  The detailed treatment of all
relevant physical processes allowed us  to investigate dynamical
properties of galaxies based on morphological classification
\citep{2015ApJ...812...29T, 2016arXiv160301619R} for the first time.
{ A small subset of these simulations also follows the evolution
of magnetic fields.}

{The web portal currently allows the user to access  a subset of the full Magneticum
simulation set. The data center hosts up} to  28 outputs of a medium size simulation
{\it Box2/hr}, which utilise $0.8\times10^{10}$ particles, covering
a volume of $(500 Mpc)^3$ as well as $11$ outputs of a larger size simulation;
{\it Box2b/hr}, which utilise $5\times10^{10}$ particles, covering
a volume of $(900 Mpc)^3$. For each cluster
contained in the simulated volumes,
the web portal { shows to the user} a 
set of pre-computed quantities. The set of pre-computed quantities is chosen to let
users select objects in categories (for example, fossils or compact objects)
that are widely studied.

\subsection{The Metadata}

\x{Each galaxy cluster object has its metadata content, that is a list
of properties associated with it (e.g. mass within a certain
radius). The metadata associated with each galaxy cluster contain table which describes its galaxy members.}

We extract
a reduced subset of halos (made of galaxies, groups or clusters, depending
on the size and resolution of the underlying simulation) by applying a
lower mass cut and providing the relevant part of the \x{available} global properties
as metadata for user queries as shown in table \ref{tab_clusters}. In addition,
for each halo we also store a list of all member
galaxies (or satellites). For each of these \x{galaxies/satellites}
we store some additional metadata as shown in table \ref{tab_galaxies}, which can
be used to further refine user queries.

We use {\sc Subfind} \citep{2001MNRAS.328..726S, 2009MNRAS.399..497D}
to define properties of haloes and their sub-haloes. {\sc Subfind} identifies substructures
as locally overdense, gravitationally bound groups of particles.
{\sc Subfind} starts with a halo list identified through
the Friends-of-Friends algorithm. For each halo and for each of its particles the local density is estimated
using adaptive kernel estimation with a prescribed
number of smoothing neighbours. Starting from isolated
density peaks, additional particles are added in sequence of decreasing
density. 
When a sequence of particles contains a saddle point in the global density field that connects two disjoint over-dense regions, the smaller
structure is treated as a substructure candidate, followed by the merging of
the two regions. All substructure candidates are subjected to an
iterative unbinding procedure with a tree-based calculation of the
potential. These structures can then be associated with galaxy clusters,
galaxy groups and galaxies and their integrated properties (like gas mass,
stellar mass or star-formation rate) can be calculated.

\subsection{Raw Simulation Data Layout}

\begin{figure}
\includegraphics[width=0.425\textwidth]{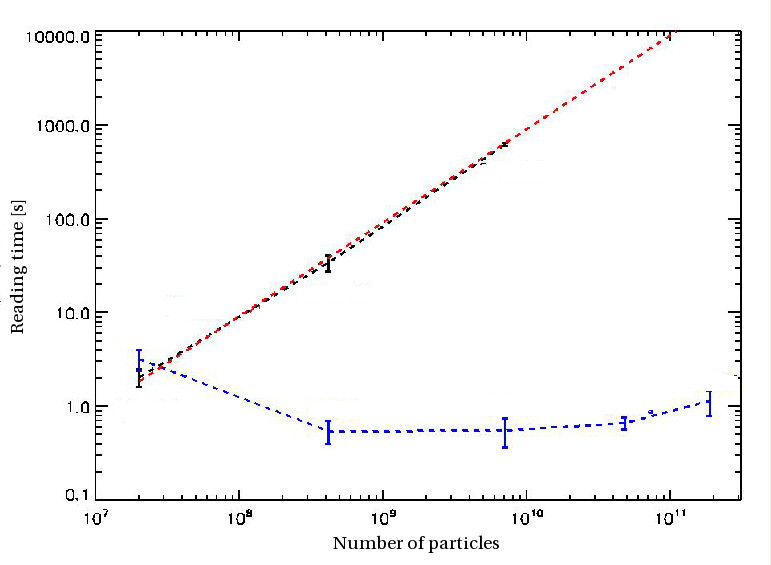}
\caption{\x{Wall clock time spent reading all data for the most
massive galaxy cluster from a snapshot { as function
of the total number of particles for increasing
simulation size}. In { black} line
there is the brute force approach by reading all data
{while the blue line} is the timing of
the improved algorithm. The improved algorithm does a spatial
selection of the snapshot by use of key-index files. Those files
allow to readout only the relevant part of the snapshot files. The
percentage at the individual data points indicate the fraction of
particles to be read compared to the overall particle number.}}
\label{fig:read_keys}
\end{figure}

{For the {\it Magneticum} project, we developed a specific output format for
very large scale, n-body, cosmological, hydrodynamical simulations. It
employs
a spatial
space filling curve to produce auxiliary data which allows a fast and direct access to
spatial regions within the output of smoothed particle hydrodynamics (SPH) simulation. This indexing scheme was extended
to work with the multi-file and multi-component output of such SPH simulations.
To optimize the access to individual data, some files contains \x{data}
regarding the content and length of individual data blocks.}

{Figure \ref{fig:read_keys} shows that the reading of all particles within the virial radius
of the most massive galaxy cluster in { any of} the simulations takes significantly less than $1$ second.
The overhead to create  and read the index list is negligible.
The algorithm speeds up also the Magneticum {\it Box0/mr} simulation post processing. This simulation utilizes
almost $2\times10^{11}$ particles and the information have to be filtered out of the individual
snapshots, which for one time instance occupy 20TB on disk. 
\ref{appendix:b} shows in detail how the index list is stored.}


\begin{figure}
\includegraphics[width=0.5\textwidth]{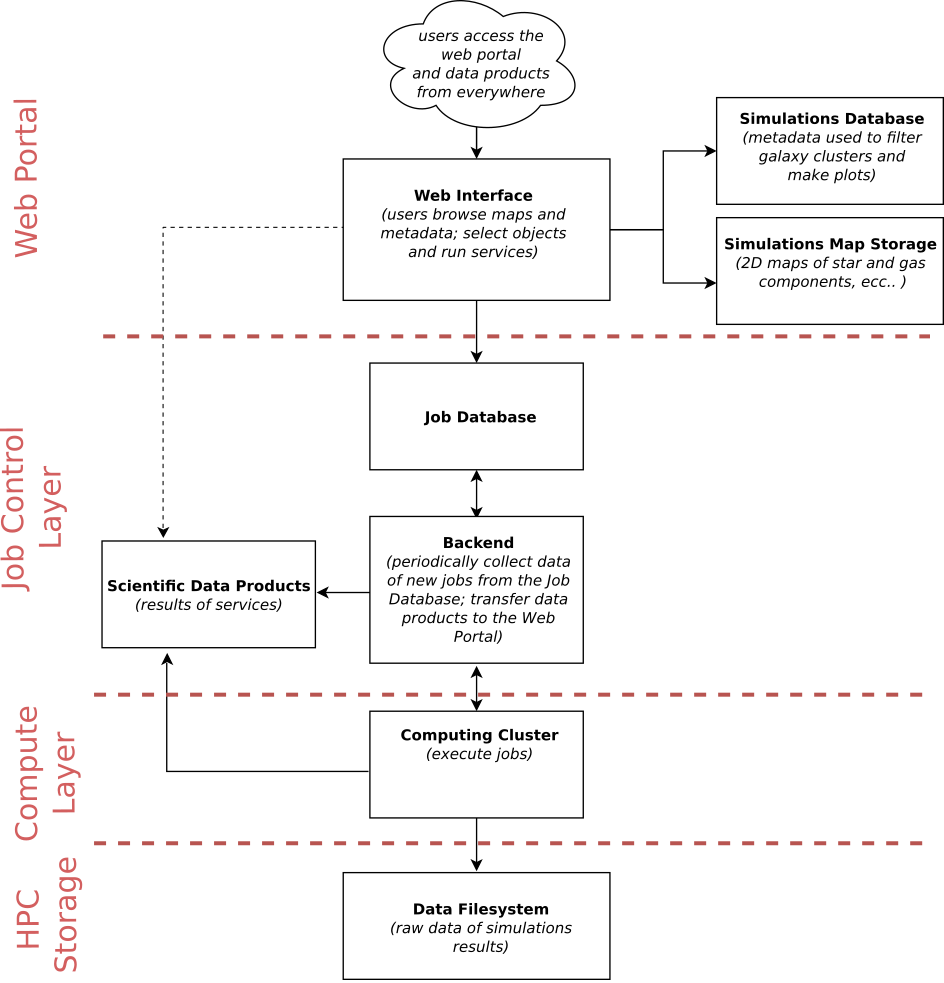}
\caption{\x{Schema summarising how the processes are distributed to different parts of the existing infrastructure,
and how the data flows are within the fully operational prototype of the web portal.} }
\label{fig:data_flow}
\end{figure}

\section{Structure of the web portal}
\label{concept}

\x{Figure \ref{fig:data_flow} illustrates our multi-layer structure
(the different layers are separated by a dashed red line).
Between those layers, data and processes flow over
the web portal, the database, the job control layer, the computing cluster
(where the analysis tools are actually executed), and the storage system
(where the raw simulation data are stored).
The need for a separation between the web interface and the backend, arises
from both the necessity of users to run personalized jobs
on raw data, managed by a job scheduler of the computing cluster and the protection
of the   data from unauthorized access.

}
\x{In \ref{appendix:c} we show how to prepare data of a simulation in order to
add it to the web portal.
}

\subsection{Overview of the multi-layer architecture}

\x{The user selects a service from the web interface and this information is
written into a job database\footnote{In our implementation we used
{\sc PostgresSQL 9.4.6} for all the different databases} (which in
our case is implemented as a separate independent instance).
The backend is { triggered by} the job database and will configure and submit jobs
to a computing cluster which will execute them.
Once a job is added, a trigger in the database will
make the backend send the job to the computing cluster.
Finally, the backend delivers the according data products to the user via a
download link which is valid for a limited time. The computing cluster
must have access to the HPC file system where the simulation data are stored,
however, it does not need to store the data locally. 
}

\x{Almost all parts are
based on common packages and available libraries except the core of the backend,
which is a customized component tailored for the data flows and job requests to
the specific needs of the scientific software performing the data analysis.}

\subsection{The Web Portal}
\begin{figure}
\includegraphics[width=0.5\textwidth]{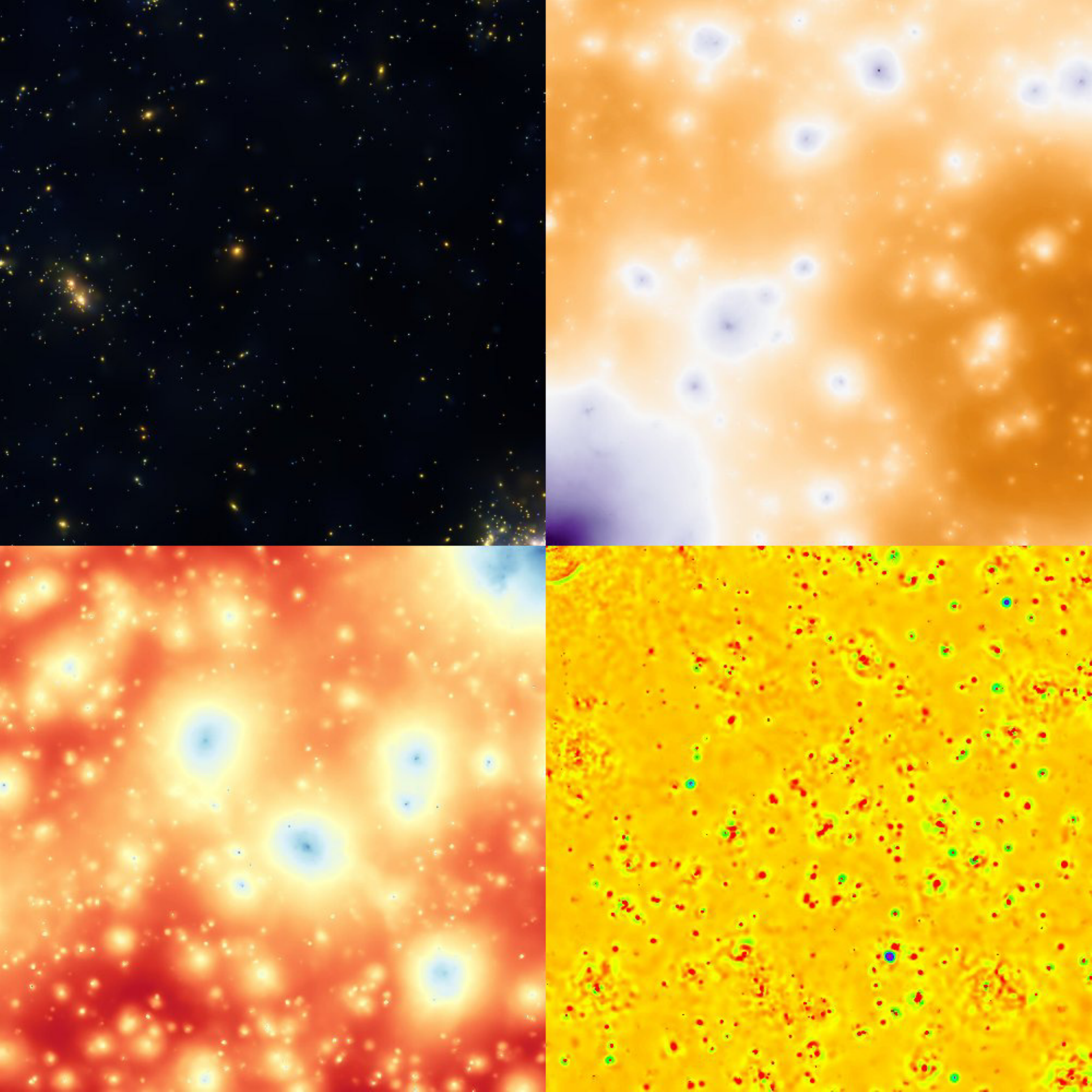}
\caption{{ The four quadrants, centred on a merging cluster,
report: the stellar component ({\it Stars}, upper left),
the ICM pressure ({\it ComptonY}, upper right), the ICM X-ray
emission ({\it ICM}, lower left) and the pressure fluctuations
({\it Shocks}, lower right).}}
\label{fig:visuals}
\end{figure}

\begin{figure*}
\includegraphics[width=0.5\textwidth]{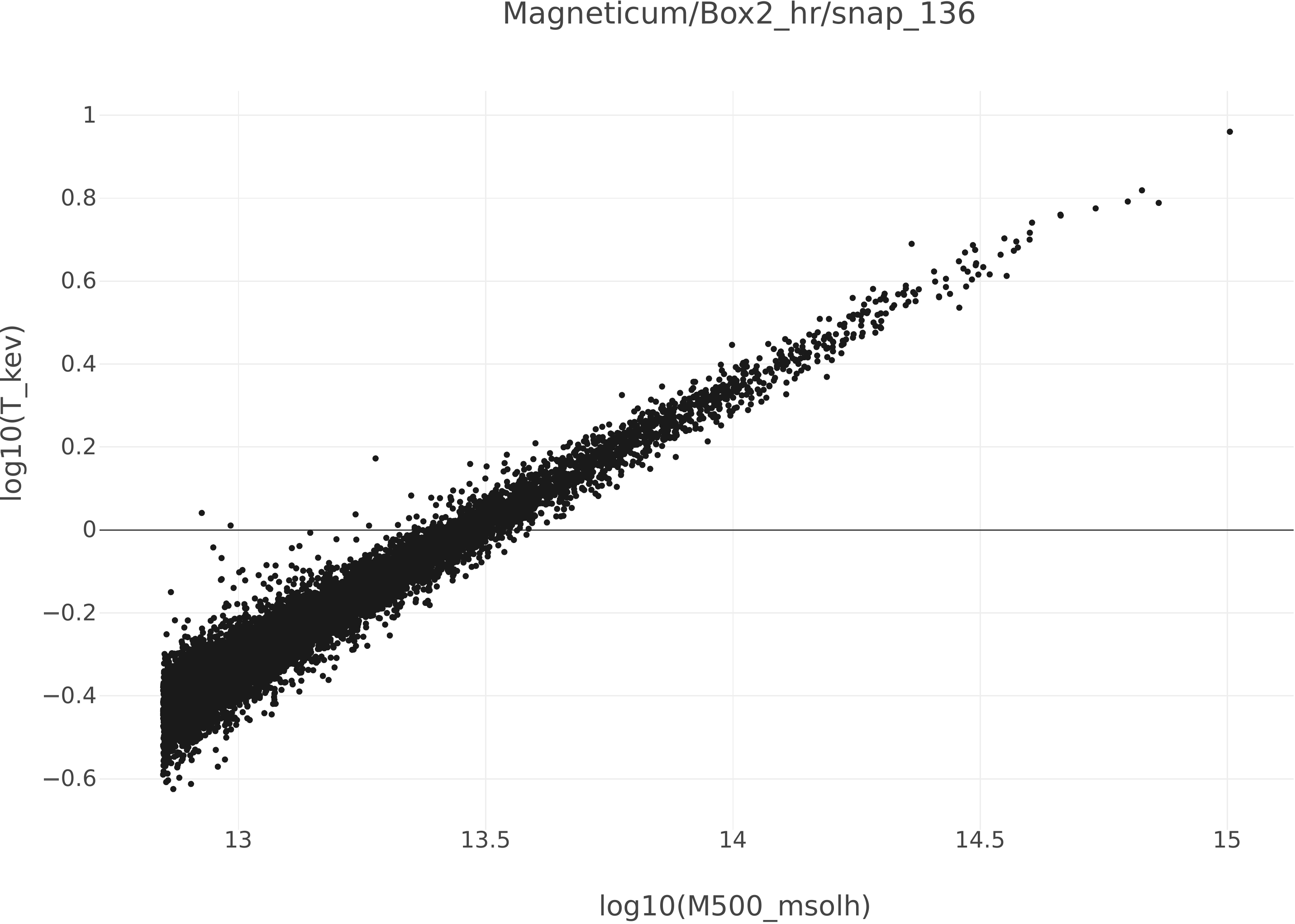}
\includegraphics[width=0.5\textwidth]{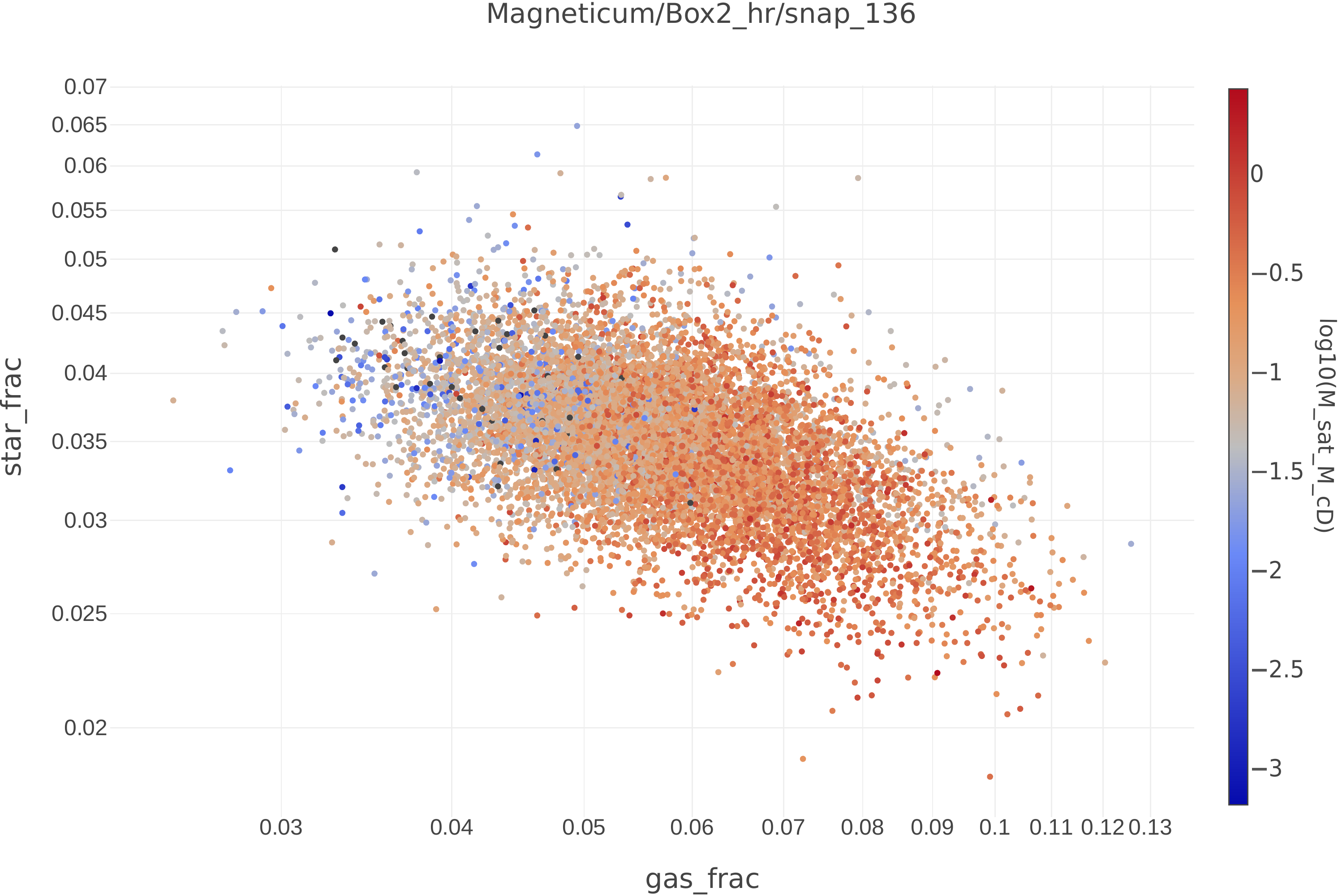}
\caption{\x{Mass-temperature correlation from the metadata of the {\it Magneticum/Box2/hr} simulation
at $z=0.14$ (left panel), where
outliers
can be identified by a red circle. The right panel shows the anti correlation of stellar and gas mass fraction, coloured by the stellar mass ratio of satellite galaxies
to central galaxy, which often is used as indicator for the dynamical state.}}
\label{fig:meta_data}
\end{figure*}

\x{The outer layer of the diagram in Figure \ref{fig:data_flow} is the
 web portal}. \x{The web portal consists of two main parts, the web interface\footnote{Our web server is built using the Python micro framework
{\sc Flask 0.11}} and the
simulations database. While this database hosts the metadata of all the simulations,
the web interface allows users to select objects in various ways through a graphical
interface. This interface
supports the visualization of  {pre-computed 2D maps rendered with Javascript}\footnote{We use
 {\sc OpenLayers 3} JavaScript library}.  Users
can navigate through different maps, as shown in figure \ref{fig:visuals} by
scrolling and zooming.
They can perform complex queries on simulation metadata based on object definitions
(in our case based on {\sc Subfind}) which are stored in a database.
}The {\sc ClusterFind} tool allows users to obtain all properties of 
clusters fulfilling the restriction criteria of the query mask.
It is then possible to download these results as
CSV-tables or to interactively visualize them. The user can thereby
make scatter plots, 1D and 2D histograms on arbitrary expressions of
the properties within the table.  In order to support the selection of interesting cluster, the user can either
select an object from the table or directly from the
scatter plots by clicking on the corresponding data point.
An example of a scatter plot between the mass
and temperature is given in figure \ref{fig:meta_data} where the
user for example could select the cluster that clearly lies outside
the mass-temperature relation.

{ Currently the web portal supports four services: {\sc ClusterInspect}, {\sc SimCut}, {\sc Smac}
and {\sc Phox}, which are described in more detail in section \ref{services}.}

The web interface can check the status of submitted jobs for
the status of a submitted job and return results to
the user.

Once the job is finished,
the backend provides the web interface with a web link, where the
user can download the results. The user is able to check the status of
all \x{jobs} and to download results via the web interface.

 \x{Metadata} from galaxies and galaxy clusters
(identified with {\sc Subfind}) are stored in a database on the web
portal. Maps can be additionally overlaid.  The data center is able to host
and handle large simulations, with order of $10^5$ objects, each
containing a number of  galaxy members. Note that a large galaxy
cluster in the simulation can have thousands of individual galaxies as
associated data.  This results in a huge number of rows in the
database.  Therefore, to achieve high performance, we used the so called
Database Indexing.  Database Indexing is a solution provided by
databases for speeding up queries on read-only tables.  In details,
Database Indexing creates a tree structure over the field of a table
so that for every indexed field there is a look-up table where data
are ordered according to this field. This speeds up every query where
data are selected on a specific range of values. {We also use a
small storage system directly on the web portal in order to store  2D maps of the
simulations.} The end user can therefore browse  quickly through these
vast simulation maps and visually seek for objects of interest.

\subsection{Job Control Layer}

The backend is based on Python and is activated when a new job is sent to the database or
when the job scheduling system starts/finishes a job.
New jobs are set up by preparing an unique job directory and template scripts and parameter
files for the different services are used. The job is submitted
to the queue of the computing cluster and executed on the cluster.
Once the job is completed, the backend will collect the results and
communicate a link for downloading the data to the web interface via
the shared job database. {The job database is connected with
both the web portal and the backend, as can be seen in
figure \ref{fig:data_flow}}. { As mentioned before,} for security reasons, the web interface can
not directly send jobs to the computing cluster or access the raw simulation
data.

\subsection{Compute Layer and HPC storage}

Large, hydrodynamical cosmological simulations nowadays are performed
within large, collaborative efforts. {While typically a large number of
scientists contributes to the development of  the underlying
code and its auxiliary functionality needed to perform specific
simulations, the actual execution of individual simulations is typically performed
by a small number of directly involved scientists.}
It is practically impossible to grant all scientists
of the collaboration (or beyond) direct access to the simulation data
on the HPC facility. Therefore we follow a different approach. While the
storage file system is typically assigned to the individual scientists
within their HPC projects, they give reading permission of the raw
simulation data to a single user of the super computing center.
This can be done, as in our case, even on a much smaller, dedicated computing cluster
where HPC storage system is available on the computing nodes. The
execution of such jobs will be typically done via a batch system.
\x{An additional, independent,} local storage keeps the data products of the {analysis tools}.
This local storage is represented by the box {\em Scientific Data Products}
in figure \ref{fig:data_flow}. The data products of the \x{services} will be
made available for the end user on the  local storage, \x{and can be shared
with the scientific community without restrictions.}
Within this concept, neither the web interface
nor the actual user ever has (or needs) any direct access to the raw
simulation data.

\subsection{Implementation}

\x{Currently,
the full outcome of the two Magneticum simulations
is stored on a Big Data storage attached to a HPC system
(in our case, it is a project space at the Leibniz
Rechenzentrum). Analysis jobs run on a separate, much smaller
scale computing
system. Specifically, we employ to this purpose
the computing cluster
at the Computational Center for Particle and Astrophysics
(C$^2$PAP)\footnote{\href{http://www.universe-cluster.de/c2pap/}{http://www.universe-cluster.de/c2pap/}}.}

\x{The web interface
is running on virtual
machines hosted at LRZ \footnote{\linka{https://www.lrz.de/services/serverbetrieb/}}.
The database of metadata properties (cluster and galaxy properties)
runs on the same virtual
machines
of the web interface. Users log in
and register by using their e-mail addresses.
User
registration, encryption and
reset of passwords are handled by the Flask-Login python
module\footnote{\linka{https://flask-login.readthedocs.io/en/latest/}}. }

Users must register and be approved by administrators of the data center to access its data.  Users register
to the web portal using their email address as username inside the
web interface. The administrator can then grant {\em roles} to the
individual users. Although at the moment all services are opened for
all registered users, the implementation allows the administrator to make
services only accessible for users with special roles.
Moreover, there is the need to share data products with a wider
community, where members do not necessarily have access to the system.
When a job is created through the web interface by a registered user,
a link containing a unique identifier of the job is delivered to the
user. \x{As mentioned above,} the link can be shared, and
its access does not require registration.

\x{To allow a first exploration of the simulations data, there is the possibility
to visually explore the simulations without being registered. However,
the access to the meta data are limited and all services are disabled
in the public browsing mode.}

Jobs that are submitted trough the web
interface are all sent to the computing cluster.

Data products are stored on the additional FTP server which runs on the 
C$^2$PAP computing cluster. Data products are guaranteed to
be stored for 30 days. The data portal and its infra structure is
 will be available for a minimum of five years.


\begin{figure}
\includegraphics[width=0.5\textwidth]{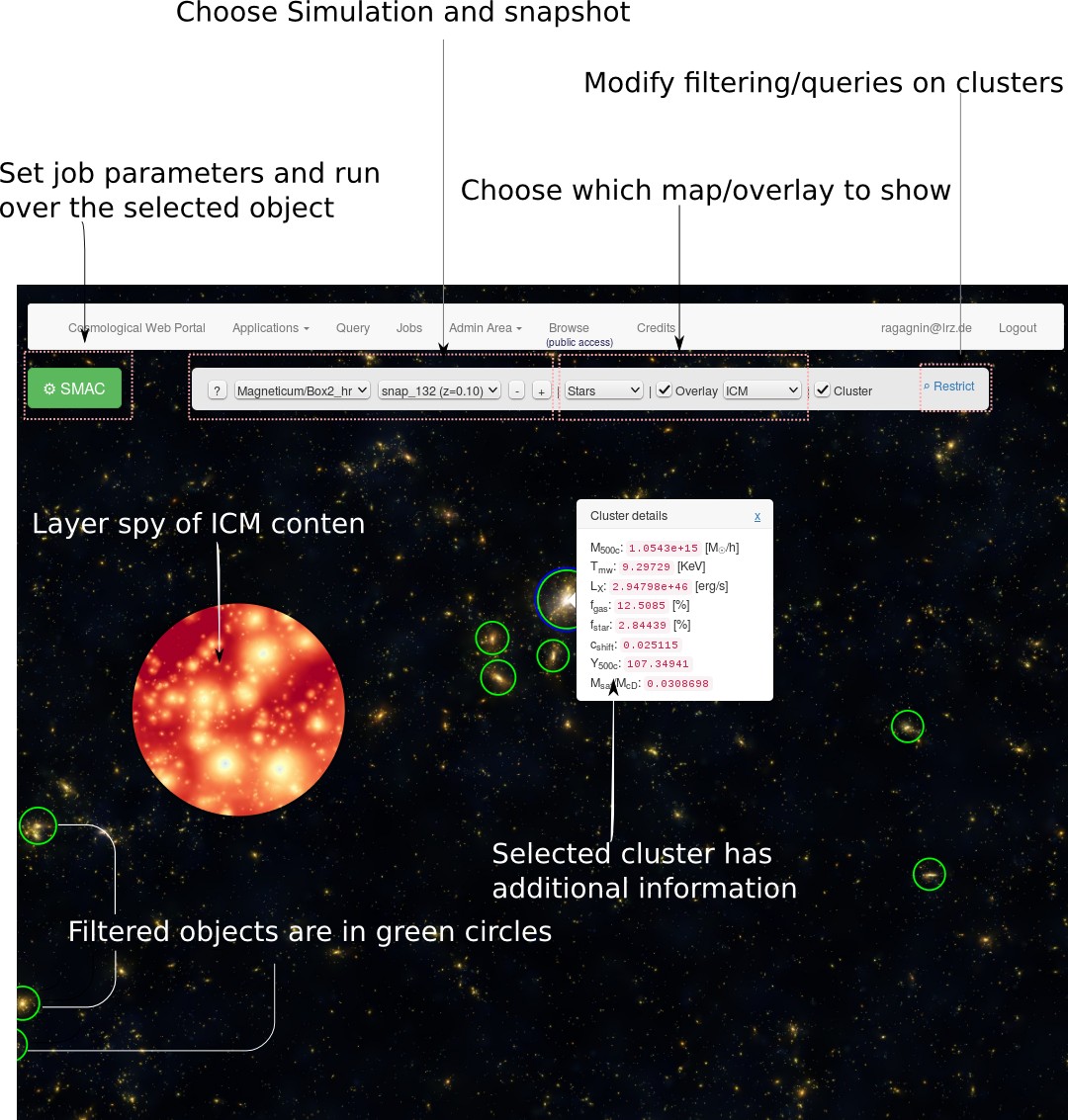}
\caption{\x{Overview of a given snapshot of a simulation. {The layer spy of ICM} can be activated within a small region around the cursor. }}
\label{fig:web_view}
\end{figure}

\begin{figure}
\includegraphics[width=0.5\textwidth]{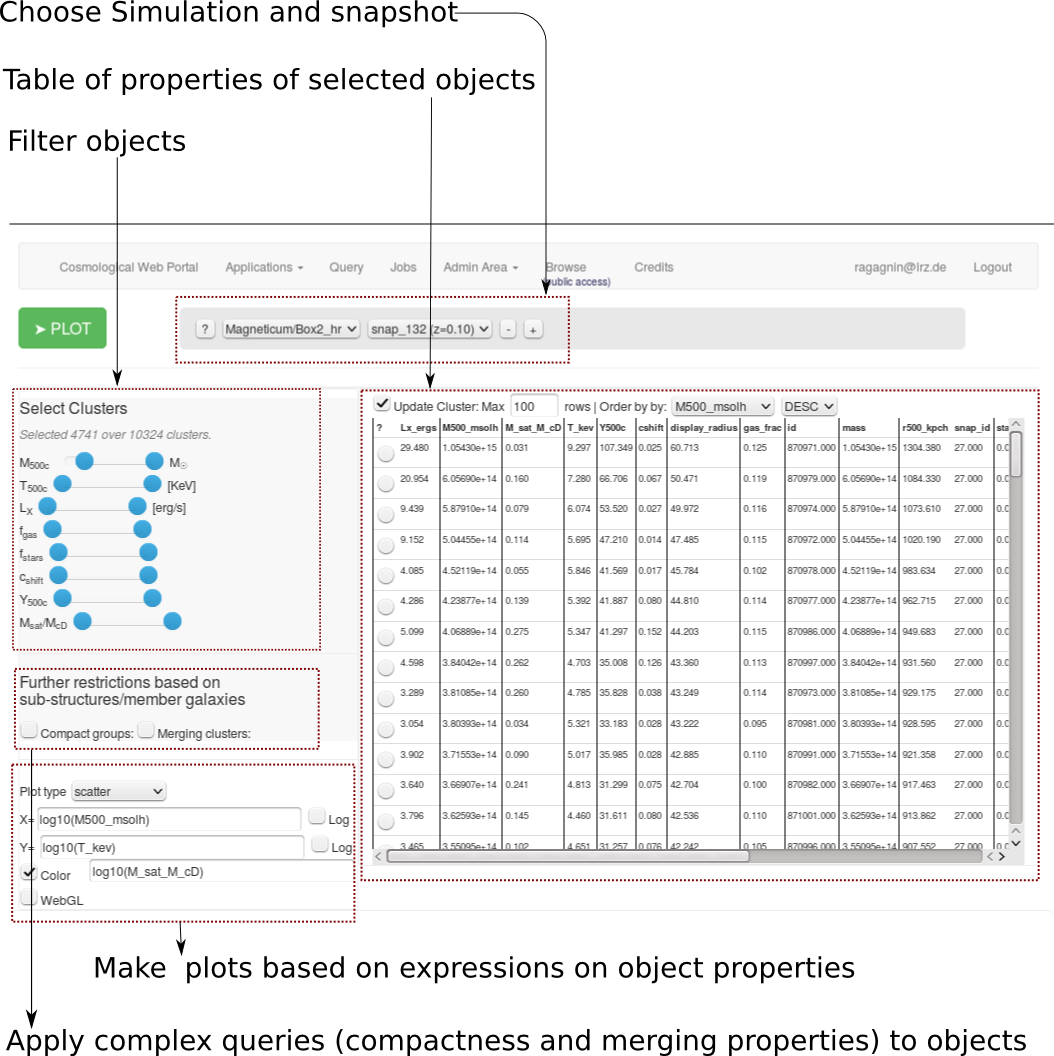}
\caption{{ A} screenshot of the service {\sc ClusterFind}. Here there is the selection panel for galaxy clusters and groups, where the objects can be filtered  according to their properties.}
\label{fig:selection}
\end{figure}

\section{Exploring Simulations}
\label{web}

Here we describe in detail how the web portal\footnote{
The web portal is a dynamic website and the interface is built using {\sc AngularJS v1.4.5},
{\sc jQuery v1.11.3}, {\sc jQuery UI v1.11.3} and {\sc Bootstrap v3.3.4}}
allows  users to  explore the simulations
and how objects of interest can be interactively found by performing complex queries \x{using {\it Restrict}
or plot the metadata quantities using {\it ClusterFind}.}

\subsection{Selecting Objects by browsing maps}

The web interface allows users to explore cosmological structures
within the simulations by panning and zooming through  high
resolution \x{ (256 mega pixel)  images}. Once a simulation is chosen,
the output at various points of the cosmological history can be
selected.
Depending on the underlying simulation, typically
between 10 and 40 different output times can be chosen. Generally, four
different maps can be selected as the prime visualization, as shown \x{already} in figure \ref{fig:visuals}.
The diffuse baryonic medium is either visualized colour coded by
it's X-ray emission ({\it ICM}) or by its pressure ({\it ComptonY}) using
{\sc Smac} \citep{dolag2005}. The stellar component is visualized
by the density of  stars and colour coded by the average age
of the stellar population using {\sc Splotch} \citep{Dolag08}. Additionally,
we computed a filtered visualization of the ICM pressure ({\it Shocks}), where shocks and
turbulence are visible. In { figure \ref{fig:visuals}
each of the {quadrants} shows one of the visuals mentioned before and demonstrates the different
appearance of the massive galaxy cluster in the centre. The arc-let (depicted by the green
curved line over the yellow background) appearing near the centre
in the {\it Shocks} visual (lower right segment) resembles a shock wave 
 which spans more than 1Mpc in size and indicates a merging cluster.}
All visualizations are based on 16k$\times$16k pixel sized images which can be
explored using the zoomify technique. In addition, the layer-spy can be activated
to switch to a different visualization view within a small region around the cursor,
(see figure \ref{fig:web_view}). This immediately gives a visual impression of the
dynamics and composition of the diffuse gaseous  and the stellar
component within our universe. It also allows users to instantaneously see
various, physical  features of the simulation. For example,
galaxies  in less dense environment appear more often in a
bluish colour, indicating a young stellar component while galaxies in
more dense environment often appear in yellow and red colours,
indicating an older stellar component.
This reflects that the underlying simulations correctly reproduce the so called
morphological density relation of galaxies, which is one of the most
prominent, observed imprints of the large scale structure of galaxy
formation.
Additionally, the position of galaxy clusters and groups can
be overlaid as green circles and an information panel on the cluster
properties is visible as soon as a  galaxy cluster is selected.

\subsection{Composing Queries}
\label{sec_selection}

To select  galaxy clusters and groups, the {\it Restrict}
window can be used to perform complex queries among the metadata of
clusters and groups, as shown in Figure \ref{fig:selection}. Every
time a value is modified by the sliders, a new database query is
performed and the selected objects are shown as green circles
in the web interface. 

\subsubsection{Restrict sample by Value}

The upper row of sliders allows users to

choose minimum and
maximum value for various global quantities, like mass ($M_{500c}$),
temperature ($T$), bolometric X-ray luminosity ($L_x$), gas and
stellar fractions ($f_{gas}$ and $f_{stars}$ respectively). The
results  can be displayed or  downloaded as ASCII tables.
This allows the user to perform simple analyses, for example plotting
scaling  relations like the well known mass-temperature relation (shown in  
the left panel of figure \ref{fig:meta_data}).
\x{For example,  a prominent outlier in the figure can then be selected in the}
web interface by restricting $M_{500c} > 14.1$ and $T > 3.05$. The
resulting cluster will then be the only marked cluster.
By a closer inspection, it can be recognized as a major merger
system, already visible in both visualizations, as double emission
peak in the diffuse medium as well as second, very large infalling
group of galaxies in the stellar component.

\subsubsection{Select by Dynamical State}

There are two classical measures \x{\citep{2013AstRv...8a..40R,2003MNRAS.343..627J,2011MNRAS.416.2997C}} of the dynamical state of galaxy
clusters and groups. One is the ratio of the total stellar mass in
satellite galaxies with respect to the mass in the central galaxy.
The other is the so called centre shift, which is  a measure of the
displacement of the ``centre of emission'' of the diffuse baryons
compared to the position of the potential  minimum and is typically
measured in units of the size (e.g. $R_{500c}$) of the system. These
two measurements can be used to select relaxed and unrelaxed
systems. For example, the outlier in the above example shows a very
large centre shift of $c_{shift}=0.13$ and a large stellar satellite
fraction of $M_{sat}/M_{cD}=0.7$, clearly classifying it as major
merger.  Usage of  the metadata tables allows users also to verify other
correlations, like the anti correlation between stellar mass
and gas mass fraction as shown in the right panel of figure
\ref{fig:meta_data}.
In that case we used one of the merger  indicators
to colour code the objects across the anti-correlation.

\subsubsection{Finding different Classes}
\label{sec:classes}
As seen in the examples before, the different filters allow users to
select objects with different global properties or with different  dynamical
states. However, such filters can also be used to select different
object classes.

{\textit{\bf Fossil groups} have typically a very large,
{dominant} central galaxy and only a very small amount of satellite
galaxies. Such objects can be selected via the $M_{sat}/M_{cD}$ parameter,
which is the mass ratio between the sum of all satellite galaxies and the
central galaxy. In this case, choosing a small value will select fossil
groups.}

{\textit{\bf Compact groups} are typically characterized by several galaxies
  of similar mass within a small spacial region. They can be selected by setting an integer $N$, from $1$ to $4,$ a maximum distance $R$ and the minimum value of the logarithm of the stellar mass ratio between the central galaxy and the $Nth$ galaxy (i.e.  $log_{10}(M_{cD}/M_{Nth})).$ 
In figure \ref{fig:selection}) a query with $N=4$, $R=100kpc$  and $log_{10}(M_{cD}/M_{Nth})<1$ returns $160$ compact group candidates out of $7428$ objects fulfilling this criteria at $z=0.67$ within our example
of {\it Magneticum/Box2/hr}.}

\textit{\bf Merging clusters} can be also selected by marking the corresponding
check-box in the {\it Restrict} window. This query allows to select
clusters or groups where at least one sub-structure fulfils the given
criteria. The user can select the range of stellar and gas mass content,
the relative, tangential or radial velocity and a distance of the sub
structure to the centre. To find bullet cluster like systems, one would
select a large stellar and gas mass, a large, outgoing velocity
(positive $v_r$) and a {large} distance from the centre.

\section{The Services}
\label{services}
The results of the different services typically come with different
number of files, depending on the configuration of the workflow
{ specified}
by the user. Therefore, the results are made available in form of
a tar ball, which the user can download. Additionally, the web
interface provides small iPython
examples for every available result.
\x{They can be used to have a quick look at the obtained results.}

\subsection{ClusterInspect}

{\sc ClusterInspect} allows users to browse the member galaxies of
the selected cluster. As in {\sc ClusterFind}, it is then possible to
make plots on the properties of the member galaxies\footnote{Within the {\sc ClusterFind} and {\sc ClusterInspect} service,
interactive plots are made using the JavaScript version of the library
{\sc plotly.js v1.9.0}}.

\subsection{SimCut}

\begin{table}
\centering
\begin{tabular}{|l|l|l|l|}
\hline
Block        & Type & Size & Blocks \\
\hline
POS &  FLOATN & 3   & 0,1,4,5 \\
VEL & FLOATN & 3   & 0,1,4,5 \\
ID  & LLONG & 1   & 0,1,4,5 \\
MASS& FLOAT & 1   & 0,1,4,5 \\
U   & FLOAT & 1   & 0 \\
RHO & FLOAT & 1   & 0 \\
HSML& FLOAT & 1   & 0 \\
SFR & FLOAT & 1   & 0 \\
AGE & FLOAT & 1   & 4,5 \\
BHMA& FLOAT & 1   & 5 \\
BHMD& FLOAT & 1   & 5 \\
BHPC& LONG  & 1   & 5 \\
iM  & FLOAT & 1   & 4 \\
Zs  & FLOATN& 11  & 0,4 \\
CLDX& FLOAT & 1   & 0 \\
TEMP& FLOAT & 1   & 0 \\
\hline
\end{tabular}
\caption{Data block names, type and sizes currently produced by {\sc Simcut}.}
\label{tab_simcut}
\end{table}

\x{The {\sc SimCut} service allows users to create artificial {\sc Gadget}
snapshot files.
They are produced following the same format of {\sc P-Gadget2.}\footnote{For more details about the Gadget formats and parameter file configurations, see: \href{https://wwwmpa.mpa-garching.mpg.de/gadget/users-guide.pdf}{https://wwwmpa.mpa-garching.mpg.de/gadget/users-guide.pdf}.}
The file produced by the {\sc SimCut} service contains a list of particles within a region centred on the selected galaxy cluster or the group.}
For every quantity that is stored (e.g. position, velocity, mass), these files have a so called ``block''.
Every block has a code name (respectively, {\sc POS}, {\sc VEL}) composed of a string of maximum of 4 characters and an array of the size of the number of particles.
In addition, the snapshot contains an extension to the {\sc P-Gadget2}  output files the so called {\it INFO} block.
This block contains information on data types and dimensions of the different values stored in the snapshot, as shown in table \ref{tab_simcut}.

\subsection{SMAC}
\begin{figure}
\includegraphics[width=0.5\textwidth]{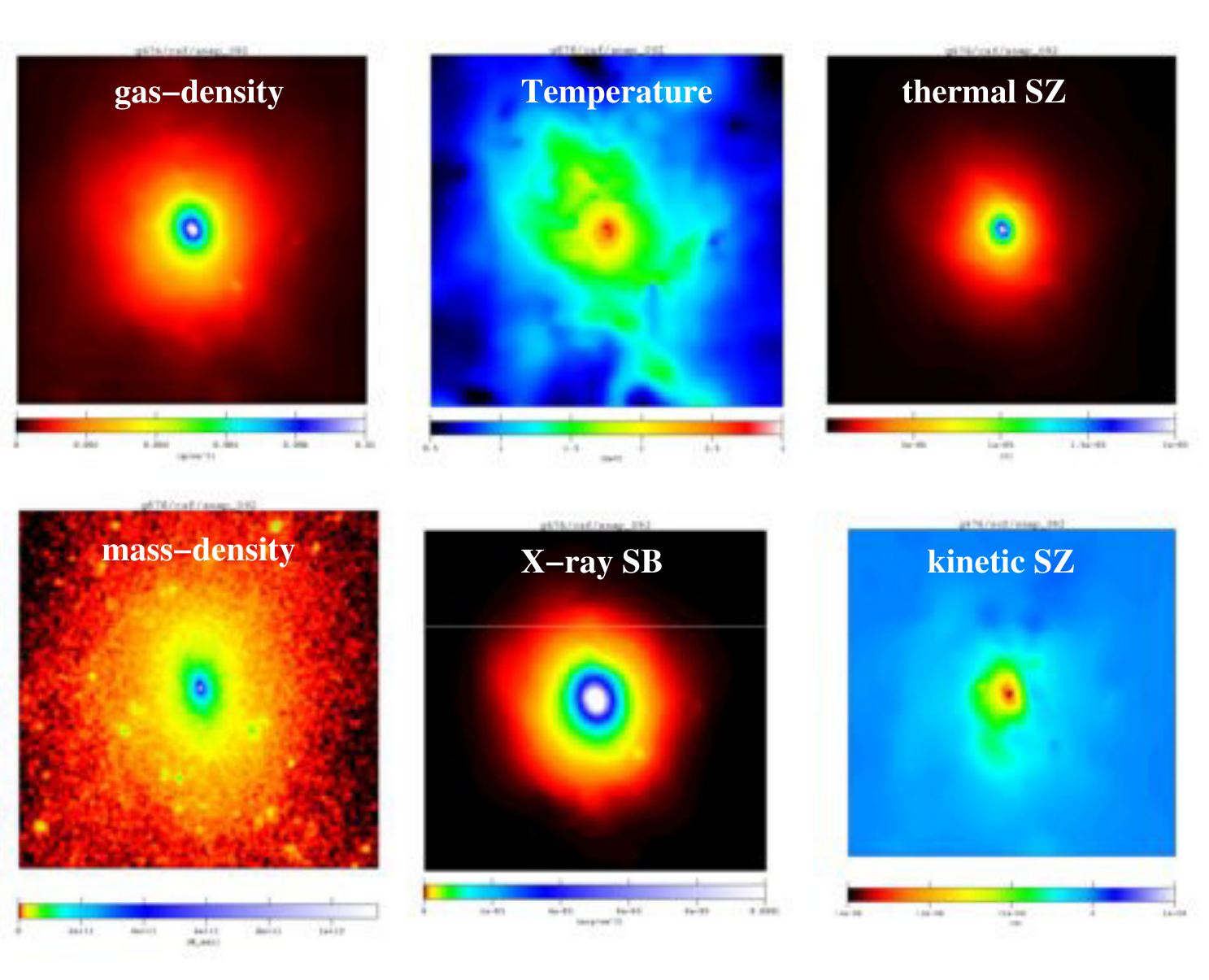}
\caption{\x{A sample of maps currently available within the the {\sc SMAC} service for a given selected cluster.}}
\label{fig:smac}
\end{figure}

The {\sc Smac} service allows users to construct maps from the simulations using the map making
program {\sc Smac} \citep{dolag2005}, which allows to integrate various physical and
observational quantities throughout the simulation. Once a galaxy cluster or group is chosen,
\x{the service allows the user to select various different map making options:}

\begin{itemize}
\item {\it Baryonic density map $[g/cm^2]$}
\item {\it Total mass density map $[M_\odot/cm^2]$}
\item {\it Mass-weighted Temperature $[keV]$}
\item {\it Bolometric, X-ray surface brightness $[erg/cm^2]$}
\item {\it thermal SZ effect [Compton Y-parameter]}
\item {\it kinetic SZ effect [Compton w-parameter]}
\end{itemize}

The size of the image as well as the integration depth along the z-axis can be chosen.
The data are returned as standard FITS\footnote{\linka{http://fits.gsfc.nasa.gov/}} files containing an image. 
Figure \ref{fig:smac} shows some example maps obtained by this service.

\begin{figure}
\includegraphics[width=0.475\textwidth]{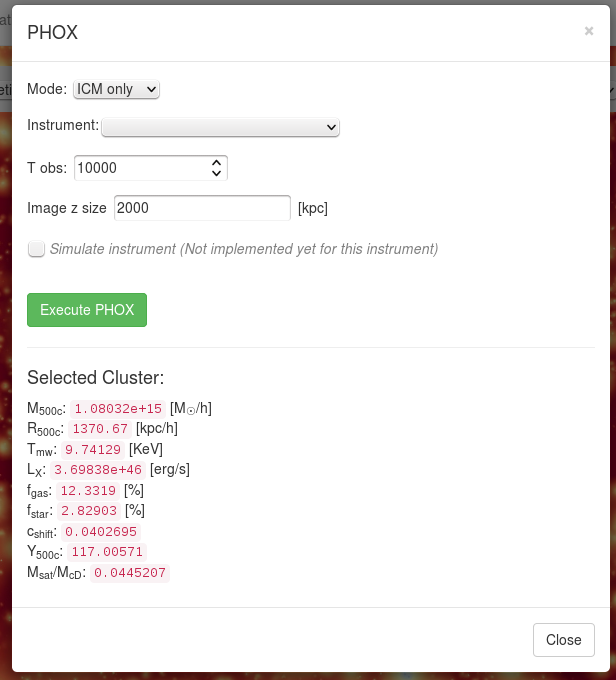}
\caption{\x{Selection panel for executing {\sc PHOX}, where details on the virtual observation
and the instrument properties can be selected}}
\label{fig:phox_select}
\end{figure}

\subsection{PHOX}
PHOX is a virtual X-ray telescope that generates X-ray synthetic 
observations from hydrodynamical, numerical simulation outputs
\citep{2012MNRAS.420.3545B,2013MNRAS.428.1395B}.
As a first step, the simulation
box is converted into a virtual box of photon packages, generated by
the sampling of the X-ray emission spectra calculated for each gas
element as well as for each AGN within the simulation.

\subsubsection{The ICM emission}
In order to compute the X-ray emission spectrum of each gas element, the
APEC model~\cite[][]{apec2001} for the emission of a collisionally-ionized,
chemically enriched plasma implemented within the
external, publicly available package XSPEC\footnote{\linka{https://heasarc.gsfc.nasa.gov/xanadu/xspec/}.}~\cite{arnaud1996}
is utilized.
The properties (namely density, temperature and
metallicity, and even chemical composition) of the gas elements are
directly used as input for this model.
Here, no external libraries of model spectra
are constructed, but the emission spectrum generated by XSPEC
is processed immediately, gas element per gas element, which
allows users to vary the { amount of} parameters involved to describe the spectral
emission without enormously increasing the computational effort and
memory requirements. Each spectrum is then populated with a predicted
number of photons (according to fiducial, general values for
collecting area and exposure time). The photons collected from all the
X-ray-emitting gas elements are eventually stored in terms of
photon packages, each of them being characterized by the position and
velocity of the original emitting element and by the array of energies
of the associated photon sample.

\subsubsection{The AGN emission}
For the AGN component (see Biffi et al. 2016, in prep.),
the procedure followed to convert the simulation box into
a box of ideal photons emitted by all the AGN sources is similar
to the one used for the gas, except for the spectral model utilized.
Namely, we model the AGN emission with an intrinsically absorbed
power law, constrained as follows.
We convert
the bolometric luminosities $L_{bol}$ of the AGN into rest-frame SXR and HXR
luminosities assuming the bolometric corrections proposed by\x{
\cite{2004MNRAS.351..169M}, which can be approximated by the
following third-degree polynomial fits
{\small
\begin{eqnarray}
\log(L_{hxr}/L_{bol})\!\!\!&=&\!\!\!-1.54\!-\!0.24\mathcal{L}\!-\!0.012\mathcal{L}^2 \!+\!0.0015\mathcal{L}^3\nonumber \\
\log(L_{sxr}/L_{bol})
\!\!\!&=&\!\!\!-1.65\!-\!0.22\mathcal{L}\!-\!0.012\mathcal{L}^2 \!+\!0.0015\mathcal{L}^3\nonumber
\end{eqnarray}
}
with $\mathcal{L}=\log(L_{bol}/L_{\odot})-12$, derived for the range $8.5 < \log(L_{bol}/L_{\odot}) < 13.$ (see fig. 3(b) in \cite{2004MNRAS.351..169M}).
Here, we mimic the}
observed scatter in these relations by adding
a random scatter of $0.1$ to the SXR and HXR luminosities,
in logarithmic scale. Then we construct an
intrinsic redshifted power law spectrum
\begin{equation}
A(E) = [K(1+z)] [E(1+z)]^{-\alpha} \left(\frac{1}{1+z}\right),
\end{equation}
where $K$ is the normalization constant, $\alpha$ is the photon index and $z$ is the redshift of the source.
For every AGN in the simulation, the two parameters $K$ and $\alpha$ can
be constrained by integrating the spectrum from the expected values of $L_{SXR}$ and $L_{HXR}$.
In this approach, the obtained distribution of photon indexes $\alpha$
reasonably reflects the observed Gaussian distribution in the range
1.2--2.8, which peaks around $\sim 2$
\citep[e.g.][]{2000ApJ...542..703Z}

Many observational works suggest that AGN sources
also show evidences for
the presence of obscuring material (i.e. the torus) in
the vicinity of the central BH,
which leads to the partial absorption of the
emitted radiation.
In order to account for this phenomenon, we assign to each AGN in the simulation a
value of the obscurer column-density ($N_h$)
by assuming the estimated
column-density distribution
presented in the study by
\citet{2014A&A...564A.125B} (see top-left panel of fig.~10, in their
paper) and based on
a sample of 350 X-ray AGN in the 4\,Ms Chandra Deep Field
South.  Within the {\sc Phox} code, we include this in the
construction procedure of the X-ray emission model from AGN
sources. The resulting absorbed SXR and HXR luminosity functions
are found to be in overall good agreement with those observed
(see Biffi et al. 2016, in prep).

\begin{figure*}
\begin{center}
\includegraphics[width=0.7\textwidth]{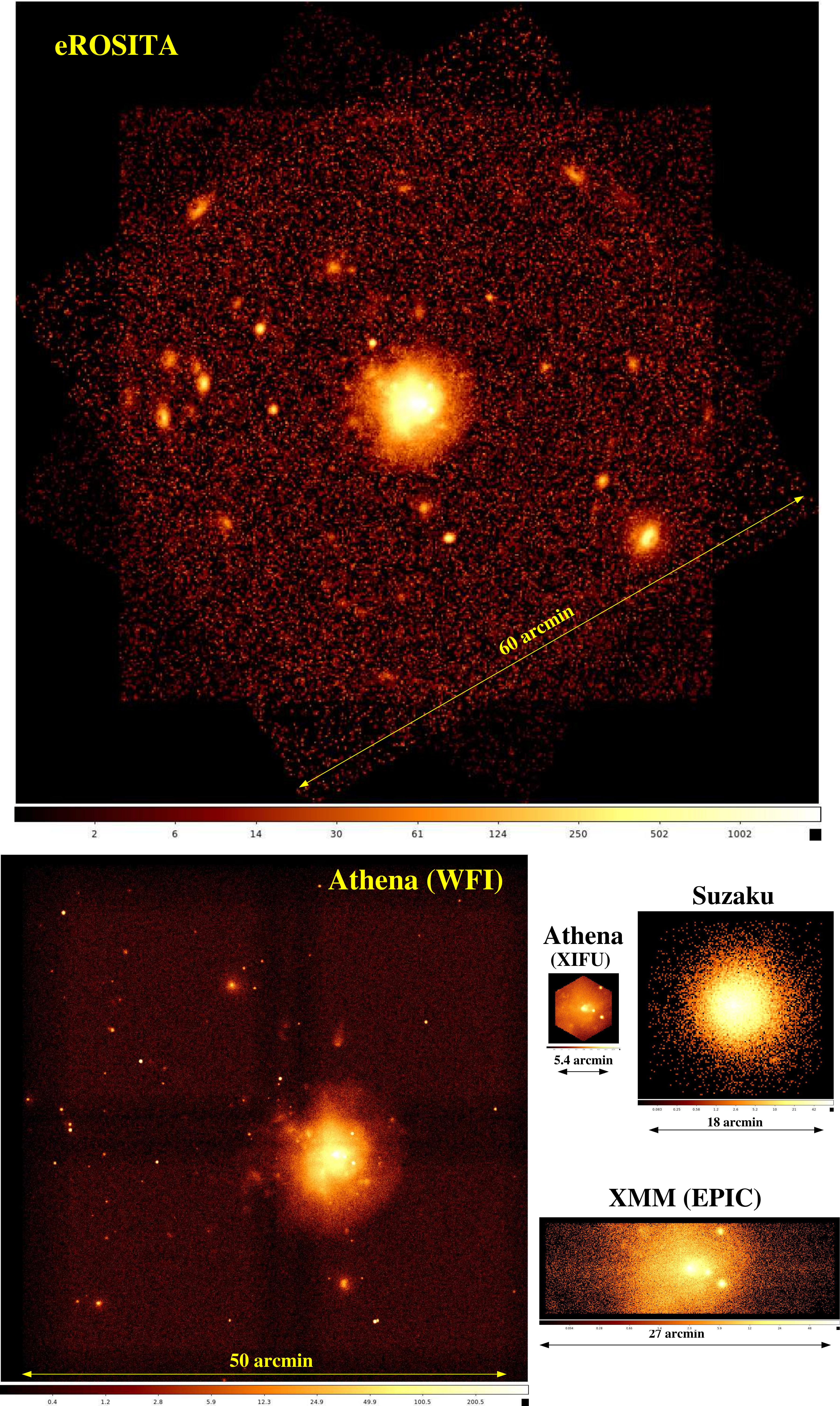}
\end{center}
\caption{Observation of the X-ray emission obtained from the combined ICM and AGN contribution,
  centred on the most massive cluster at $z=0.3$, performed with different, current and
  future X-ray instruments, including the actual instrument simulations.}
\label{fig:instruments}
\end{figure*}

\subsubsection{Performing an X-ray Observation}

With PHOX, the field of the selected photon packages (either
only ``ICM''/``AGN'' or both can be selected) is further processed
by taking into account
the geometry of the mock study and the {\it idealized} instrument
characteristics. 
In particular, given the ideal
cube of virtual photon packages associated to the simulation output,
it is here possible to select the size of the subregion of interest,
centred on the selected galaxy cluster or group,
and to choose
an instrument, which defines a field of view and a nominal
effective area of the X-ray telescope, as shown in Figure
\ref{fig:phox_select}. This also defines the line-of-sight (l.o.s.)
direction and accounts for the according Doppler shift in the photon
energies due to the peculiar motion of the original emitting
particles. The virtual observation then returns the photon list
expected for the idealized instrument chosen (assuming the
nominal effective area over all energies, no beam smearing of the
position and keeping the exact energy) and returning the
data in the form of a FITS file in {\sc Simput}\footnote{\linka{http://hea-www.harvard.edu/heasarc/formats/simput-1.1.0.pdf}.} format,
which allows users to directly { utilise} this file for
more sophisticated
instrument simulations. The time of the virtual observation is restricted, and a
limit of $A_{instrument}\times{}T_{obs} < 10^9$ is enforced. This
process typically takes only few seconds to be performed.

\subsubsection{Including the mock X-ray observation}

We also installed some publicly available instrument simulators, which
can be added by clicking on the {\it Simulate Instrument} check-box,
for some of the chosen instruments.

For {\it XMM(EPIC)}, {\it eROSITA} and {\it Athena(XIFU/WFI)}
we are using {\sc sixte}\footnote{\linka{http://www.sternwarte.uni-erlangen.de/research/sixte}.}, 
whereas for
{\it SUZAKU(XIS)} we are using {\sc xissim}, which is part of the {\sc
  HEADAS} package \citep{2007PASJ...59S.113I}. { For {\it XraySurveyor(HDXI/XCAL)}
  and {\it Hitomi (SXS/SXI/HXI)} we are using {\sc simx}\footnote{\linka{http://constellation.gsfc.nasa.gov/science/simulatorInstall.html}} while for {\it Chandra (ACIS-S/I)} instruments we are using {\sc marx} \citep{2012SPIE.8443E..1AD}.} Depending on the
instrument and  time chosen for the observation, this process can take up to several
minutes. It returns an event file which can be then analysed with
standard X-ray analysis tools.

Figure~\ref{fig:instruments} shows the
result for observations centred on the same massive galaxy cluster at $z=0.3$
while choosing different instruments. The exposure time is set to 100ks
(reduced to 50ks for {\it Athena}) and a slab of 100Mpc along the line of sight
is used. Both, the ICM as well as the AGN component are taken into account in this example.
For the {\it eROSITA} case, a simulation of all 7 CCD chips is preformed, delivering 7 event files,
which are additionally combined to one single event file, which leads to the appearance of
edges from the rotated frames in the combined image as seen in the map.
For the {\it Athena(WFI)} simulation, the 4 individual chips are simulated,
{ as still visible by the gaps in the maps. However, as for real observations,
the telescope is made to dither during the integration time to smears out the
gaps between the chips, which is controlled
by a realistic attitude file which defines the pointing of the instrument.}

\x{In \ref{appendix:a} we show configuration files for {\sc sixte}, {\sc xissim}, {\sc simx} and {\sc marx}
  used for each instrument we included.}

\section{Conclusions}

In this work we {present a data center based on a multi-layer infrastructure for}
large cosmological hydro-dynamical simulations. {It will give}
a wide scientific community
the possibility {to perform analysis tools} on data
from {several} large simulations.

The increasing {amount} of upcoming astronomical surveys { makes} it necessary
to increase  resolution and volume of cosmological simulations  to interpret the results from such observational
campaigns.
These simulations may include different matter contents 
and object types (gas, dark matter, stars, black holes) and various
{ physical} processes that shapes the formation of objects and structures in our universe.
All those factors dramatically {increase} the {output} size { of} current
simulations.

{ As a result, the storage and sharing of the largest simulations
  that are currently available represents a real challenge.}

For example, the size of an individual snapshot of the {\it Box0/mr}
simulation of the {\it Magneticum} project (which follows almost
$2\times10^{11}$ particles), is more than 20TB.

Although we currently only store results from the {\it Magneticum} project,
the data center infrastructure is capable of including other simulations as well.

This  data center allows users to configure workflows to run
individual {analysis tools} for specific objects of a given
snapshot on the raw simulation data. A web interface helps
the user to build compounded queries on metadata of the different
simulations, which in our case are galaxies and galaxy clusters as
obtained by {\sc Subfind}.
This allows  users to select a target object in the different simulations with the desired properties.
Those queries can be restrictions to global properties like mass or
other observables, as well as complex queries on various internal
aspects. This gives the possibility to select different general classes of objects (like
merging clusters, compact groups or fossil groups).
Our initial version of the  data center  provides so far
 { four services:
{\sc ClusterInspect} plots and shows data of member galaxies of a given galaxy cluster,
{\sc SimCut} gives the possibility of downloading the raw simulation data of all
particles belonging to the selected object, {\sc Smac} creates 2D
maps of different physical or observational quantities, {\sc Phox}
generates  virtual X-ray observations.}

A number of considerations about accessibility, security and
performance led us to a multi-layer infrastructure: the web portal, the job control layer,
a computing cluster and the HPC
storage system. The outer layer lets people perform highly compounded and
elaborated queries on {\sc Subfind} results; the computing
cluster {runs} the chosen job, reading the raw simulation
data directly via access to the HPC storage system and { sends}
the results  back to the web portal.

Concerning the cluster selection, there is the possibility of
interlocking a number of different subqueries. Some very basic queries
are filtering by mass, temperature, bolometric X-ray luminosity,
fraction of stars and gas masses. The user can also select
an object by browsing 2D maps of the simulation, or by clicking
objects in scatter plots of {\sc Subfind} data.

It is possible to run queries for the selection of clusters in
different dynamical {  states. For instance depending on their}
satellite fraction and displacements between baryon and potential
centre. Other subqueries allow for the selection of clusters with
different degree of compactness by choosing the ratio between masses
of the central galaxy and the n-th satellite galaxy within a given distance.

Services themselves {come} with different parameters that the user is
free to choose. {\sc Smac} can project onto  three different axes
and map different matter properties (i.e. density, temperature, X-ray
temperature); {\sc Phox} can simulate a number of different
instruments (i.e. XMM, Suzaku, Chandra, eROSITA, {Athena, XraySurveyor, Hitomi}) and compute X-ray
emission due to AGN or ICM, separately or together.

Services will be extended in the future to support additional analyses. Possibility to browse observationally motivated realizations of
light-cones {is} foreseen. Finally, the  data center  will permanently grow in
size and amount of simulation data which will be made available for the
general scientific community. Although some modifications on the different
components would be needed to adapt the concept to other infrastructures
and simulations, we are willing to provide the underlying source code on request and
to give advice for adapting the package to other institutions. {The current
infrastructure for the  data center  is secured for five years.}

\section{Credits}

The results obtained via the  data center  are free to use.
Please acknowledge the {\it Magneticum} Project\footnote{\linka{http://www.magneticum.org}} (Dolag et al. 2016, in prep)
and the {\it Cosmological Web Portal} (this work, Ragagnin et al. 2016). Please also give
the references for the individual simulations used (Box2/hr: \citet{2014MNRAS.442.2304H},
Box0/mr: \citet{2016MNRAS.456.2361B}).
If results produced with PHOX are used, please cite \cite{2012MNRAS.420.3545B}, and in case of SMAC please cite \cite{dolag2005}.
In the case that results of the X-ray instrument simulators are used, please refer to
{\sc sixte}\footnote{http://www.sternwarte.uni-erlangen.de/research/sixte},
{\sc xissim} \citep{2007PASJ...59S.113I}, 
{\sc simx}\footnote{http://constellation.gsfc.nasa.gov/science/simulatorInstall.html} 
and {\sc marx} \citep{2012SPIE.8443E..1AD},
depending on the simulator used for the simulation.

\section*{Acknowledgements}
We thank G. Lemson for various discussions and his contribution to various VIO
proto-types, which largely inspired this work.
We thank T. Dauser, G. Lamer, N. Clerc, A. Rau and E. Pointecouteau for helping with the X-ray data and
especially with the instrument simulators, which can be additionally involved within the ``PHOX'' service.
AR thanks Nathan James Turner for carefully checking the  grammar of this manuscript.
We also thank Ben Hoyle for carefully reading the manuscript.
This work has been supported by the DFG Cluster of Excellence ``Origin and Structure of the Universe''  (\href{www.universe-cluster.de}{www.universe-cluster.de})
and the SFB-Tansregio TR33 ‘The Dark Universe’.
We are especially grateful for the support by the Computational Center for Particle and Astrophysics
(C$^2$PAP).
Computations have been performed at the Leibniz Supercomputing
Centre (LRZ) with CPU time
assigned to the Project ``pr83li'' and ``pr86re''. The project received
technical support from LRZ as well. Information on the {\it Magneticum
Pathfinder} project is available at \\
\href{http://www.magneticum.org}{http://www.magneticum.org}.


\bibliography{master3}

\appendix
\section{Indexing files}
\label{appendix:b}

\begin{figure}
\includegraphics[width=0.425\textwidth]{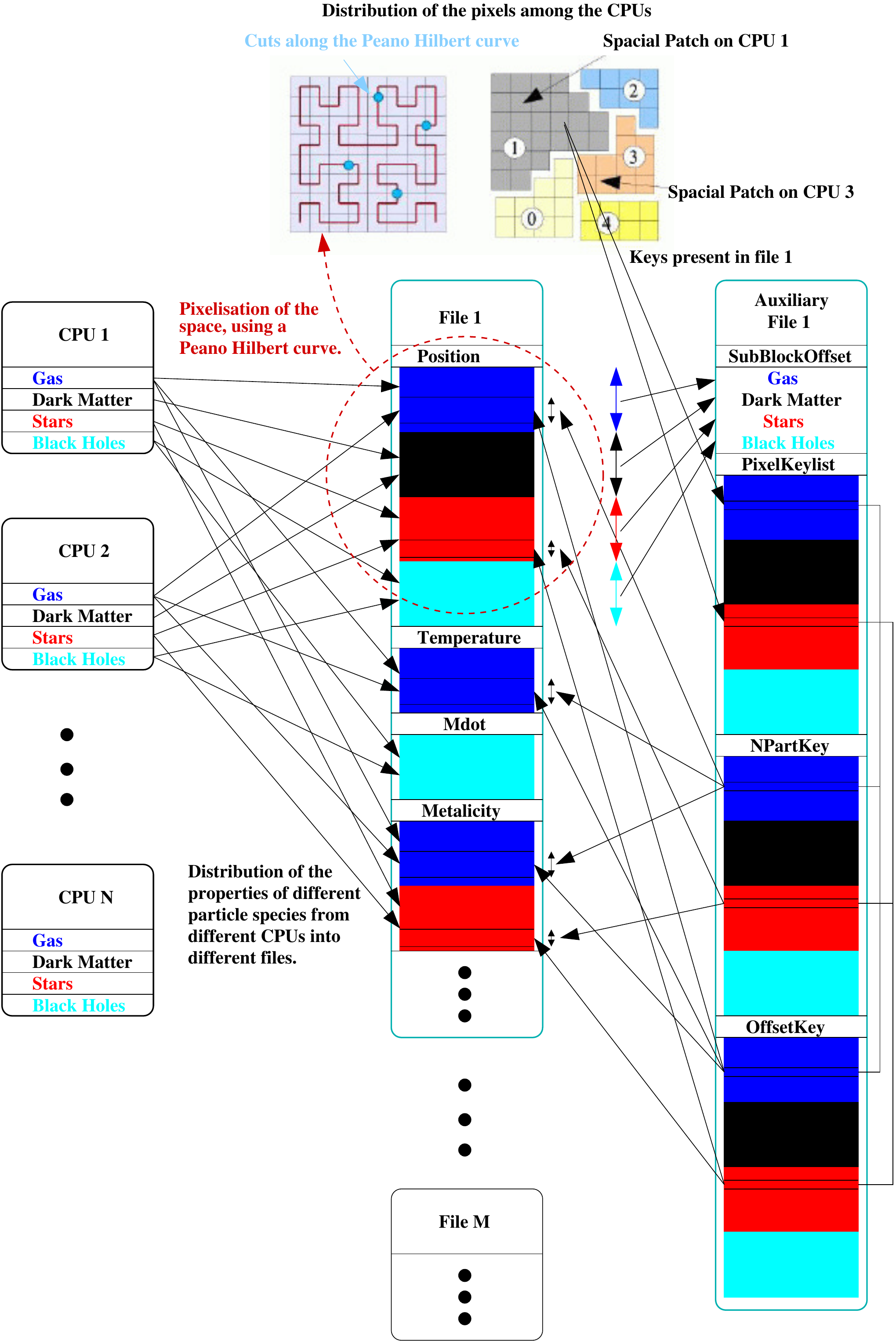}
\caption{This diagram illustrate how data of the different properties, attached
to the different particle types are collected from several CPUs and stored
into different files (left part). Once the position of the particles are
associated (and sorted) according to their associated elements along a
space filling curve, the index of such {\it pixels}, together with the
number of particles associated with each {\it pixel} and the according
offset in the files can be stored as auxiliary data. }
\label{fig:key_files}
\end{figure}

Below we explain how to build the auxiliary data files in order to add third-party simulations to the  data center.
Figure \ref{fig:key_files}
shows a sketch of how the indexing scheme works for the multi-file and multi-component output of our SPH simulations. In such cosmological SPH simulations, the data
to be stored is quite complex. The particle data are distributed among many thousands
of CPUs. Each CPU holds various  particle  types, representing
different components of the simulated system. For these particles, different
properties have to be stored. Some (like {\it temperature}) are specific to
individual particle species (like {\it gas} particles in this case). Other
properties (like {\it position}) are necessary for all components.
\x{To optimize the I/O and to avoid bottlenecks, only a subset of CPUs write in parallel.
Data from the individual CPUs are stored in individual sub-files, as illustrated in the left
part of figure \ref{fig:key_files}.}
To optimize the access to individual data,
files are structured with \x{additional data} regarding the
content and length of individual data blocks.

We implemented an algorithm which sorts the particles of all CPUs among
a coarse-grained space filling curve before writing them.
In addition it produces an auxiliary file which allows \x{identification}
of the sub-data volume elements of any stored property among all particle
species associated with each element of the space filling curve.
This allows the user to effectively collect all data associated with a given volume
in space introducing minimal overhead.
\x{In detail, the process works as follows:}

\begin{itemize}
\item The particles will be ordered along a space filling curve (in our case, a  Peano
Hilbert curve) with a defined graininess (i.e. {\it pixel} size) \footnote{
Note that in contrast to the standard domain decomposition in Gadget
which is using a 64 bit long key, we are using a shorter, 32 bit long key
which strongly reduces the later reading overhead.} among all
CPUs. This order will be used to decompose the spatial region of the
simulation among the different CPUs so that all particles
within an individual {\it pixel} are associated to a
CPU. During this process, the amount of particles of different species
falling within each {\it pixel} is stored as well.
\item The particle data of the various CPUs are written into the
sub-files in the same order as above.
Therefore particles belonging to the same {\it pixel} are written in a  consecutive order into the files.
\item \x{ An additional auxiliary file   stores the
information needed to re-construct the position of each sub-data-block
for particles located in the a specific {\it pixel}}.
It contains the offsets between different particles species, the
list of {\it pixels} present in the file, the number of particles in the
\x{related} {\it pixel} and the offset of each {\it pixel} within each particle
species.
\item Finally a super index file is created, which contains the compressed information
of {\it pixel} indexes and the files where they are contained.
This super index is later used to recover which sub-files have to be accessed
for the reading process.
\end{itemize}

The  additional auxiliary files are produced for each output of the simulations and
allow the user to investigate very efficiently individual objects within such cosmological
simulation through post-processing tasks. The post-processing software has been adapted
to this new output and for reading sub-volumes of the simulation. The following strategy
is applied for a given sub-volume of the simulation which should be analysed
(e.g. given by a galaxy cluster position and its size):

\begin{itemize}
\item First,  a list of
{\it pixels} (i.e. the elements of the space filling curve which falls within
the region of the space of interest) is prepared.
\item Then the super index file is read and a list of the output files
which hold the individual subsets of the {\it pixel} index list is produced.
\item For each sub-file in that list, first the  {\it pixel} list is read
from the auxiliary file and then it is compared with the {\it pixels} needed for the current
task. For this subset of pixel, the  additional information
 is read from the auxiliary file.
\item Now the post-processing can reconstruct the position
of the sub-blocks which have to be read from the individual blocks
within the data file. This sub-blocks can then be read directly without
loading the full data block. As further optimization, consecutive
sub-blocks within individual files, are joined to larger sub-blocks to prevent
unnecessary fragmentation of the reading process.
\end{itemize}

\section{Preparing Simulations for the  data center }
\label{appendix:c}
\x{In this appendix we describe
how to add simulations to the data centre. To this scope, several pre-computed
data products, meta data and informations have
to be provided, as we briefly describe in the following.}

\subsection{Describing the Simulations}

\x{Every simulation consists of one or more time instances (snapshots),
where each one should consist of a set of pre-computed 2D maps in the
{\em zoomify}\footnote{http://www.zoomify.com/free.htm} format, a
list of galaxy clusters and their galaxy members. Metadata regarding
galaxy cluster properties and member properties are stored in
{\em yaml\footnote{see \href{http://yaml.org/}{http://yaml.org/}}} files.
Metadata and maps are stored in files and folder in a way that for every
simulation there is a folder for its 2D maps ({\em zoomify} folder)
and one for the metadata ({\em simulations} folder). Here follows a
sample of the folder structure:}

\dirtree{%
.1 /.
.1 zoomfy.
.2 \$SIMULATION\_NAME.
.3 \$VISUAL1\_SNAPNUM.
.4 TileGroup0.
.4 TileGroup1.
.1 simulations.
.2 \$SIMULATION\_NAME.
.3 meta\.yml.
.3 \$SNAPNUM.
.4 meta.yml.
.4 cluster.txt.
.4 galaxies.txt.
}

\x{The {\em zoomify} files are stored on the local disk space of the web portal
and contain the images which can be browsed by the web interface.
Table \ref{tab_simulation} shows a list of parameters used to describe a simulation.}

\begin{table}
\centering
\begin{tabular}{|r|l|}
\hline
Property       & Description \\
\hline
name         & Simulation name \\
code         & Description of the Code used \\
size\_mph     & Mpc/h \\
n\_particles  & number of initial particles \\
nfiles       & number of files the snapshot \\
mdm\_msun     & mass dark matter particles [Msun/h] \\
mgas\_msun    & mass gas particle [Msun/h] \\
epsilon\_dm   & softening of dark matter particles \\
epsilon\_gas  & softening of gas particles \\
epsilon\_stars & softening of stars particles \\
f\_bary      & cosmological baryon fraction \\
omega\_m     & total matter content \\
omega\_lambda & Cosmological constant  \\
hubble       & Hubble constant \\
sigma\_8      & Normalization of matter power-spectrum \\
n\_primordial & Primordial power spectrum index  \\
path\_exec    & file system path to simulation \\
\hline
\end{tabular}
\caption{Properties used to describe a simulation.}
\label{tab_simulation}
\end{table}

Every time instance of a simulation, which is \x{stored in the  data center} and made
available via the web interface, contains its own sub directory which
holds the prepared metadata for clusters and galaxies. They contain
the \x{information} as described in the previous section. In addition, it
also contains a general {\em yaml} metadata file with snapshot details.
Table \ref{tab_snapshot} contains the parameters that
typically describe a snapshot.

\begin{table}
\centering
\begin{tabular}{|r|l|}
\hline
Property       & Description \\
\hline
name &               snapshot name\\
mask\_path &          not used\\
width\_pixel &       total number of pixels\\
height\_pixel &      total number of pixels\\
redshift &          Redshift \\
angular\_diameter &  Angular diameter distance\\
phox\_max &          Maximum GRASP (for Phox)\\
phox\_avail &        Phox service available\\
smac\_avail &        Smac service available\\
simcut\_avail &      SimCut service available\\
simcut\_plot &       SimPlot service available\\
\hline
\end{tabular}
\caption{Properties used to describe a snapshot within a simulation.}
\label{tab_snapshot}
\end{table}

Such information and definitions can be extended easily
to any other cosmological simulation and will allow the final user to flexibly
add new 
simulations
to the system. Also, services can be
readily
added or disabled for individual simulations.

\begin{table}
\centering
\begin{tabular}{|c|c|c|}
\hline
quantity       & unit            & name     \\
ID             & [integer]       & \texttt{id}  \\
$x$            & [kpc/h]         & \texttt{x} \\
$y$            & [kpc/h]         & \texttt{y}  \\
$z$            & [kpc/h]         & \texttt{z} \\
$M_{500c}$      & [M$_\odot$/h]   & \texttt{M500\_msolh}   \\
$R_{500c}$      & [kpc/h]         & \texttt{r500\_kpch} \\
$f_{gas}$       & [fraction]      & \texttt{gas\_frac} \\
$f_{stars}$     & [fraction]      & \texttt{star\_frac}  \\
$L_x$          & [$10^{44}$erg/s] & \texttt{Lx\_ergs} \\
$Y_{500c}$     & [$\Delta{}T/T$]  & \texttt{Y500c} \\
$M_{sat}/M_{cD}$& [fraction]      & \texttt{M\_sat\_M\_cD}  \\
$c_{center}$    & [$R_{500c}$]    & \texttt{c\_shift}  \\
\hline
\end{tabular}
\caption{Metadata for galaxy clusters and groups, from top to bottom:
cluster ID, position (in comoving coordinates), mass and radius in
respect to 500 times  the critical density, gas and star fraction,
bolometric X-ray luminosity, stellar mass fraction of satellite
galaxies to central galaxy,  \x{and weighted centre shift between
X-ray emission and mass distribution within $R_{500c}$.}}
\label{tab_clusters}
\end{table}

\begin{table}
\centering
\begin{tabular}{|c|c|c|}
\hline
quantity            & unit            &  name     \\
ID                  & [integer]       & \texttt{id} \\
$x$                 & [kpc/h]         & \texttt{x}  \\
$y$                 & [kpc/h]         & \texttt{y} \\
$z$                 & [kpc/h]         & \texttt{z}  \\
$M_{star}$          & [M$_\odot$/h]   & \texttt{M\_solh} \\
$M_{gas}$            & [M$_\odot$/h]   & \texttt{M\_gas} \\
sfr                 & [M$_\odot$/year] & \texttt{sfr\_msoly}  \\
host ID             & [integer]       & \texttt{cluster\_id} \\
radial distance     & [kpc/h]         & \texttt{dist}          \\
\x{peculiar} $v_x$               & [km/s]         & \texttt{vx}  \\
\x{peculiar} $v_y$               & [km/s]         & \texttt{vy} \\
\x{peculiar} $v_z$               & [km/s]         & \texttt{vz}  \\
velocity   & [km/s]          & \texttt{dv}          \\
radial velocity     & [km/s]          & \texttt{dr}          \\
tangential velocity & [km/s]          & \texttt{dt}          \\
mass ratio to cD    & [km/s]          & \texttt{log10\_mcD\_m}   \\

\hline
\end{tabular}
\caption{\x{Metadata} for galaxies, from top to bottom: galaxy ID,
position (in comoving coordinates), mass, star-formation rate, ID of
cluster or group where the galaxy belongs to, distance to the
\x{center} of the cluster or group it belongs to, \x{different velocity
components relative to the cluster center (as peculiar velocities) and
the ratio of the stellar mass in satellite galaxies in respect to the
central one.}}
\label{tab_galaxies}
\end{table}

\section{Instrument Configurations}
\label{appendix:a}

{ In this Appendix we report example configuration files for {\sc xissim}, {\sc
    sixte}, {\sc sim} and {\sc marx}, used for each instrument currently included
    within the {\sc Phox} service. Note that the exposure time will be replaced by 
    the value chosen in the web interface.}

{\it Suzaku(XIS):}
\begin{lstlisting}
xissim \
  clobber=yes \
  instrume="XIS1" \
  ea1=0 ea2=0 ea3=90 \
  infile1="phlist_xissim.fits" \
  infile2=none \
  date_obs ="2009-09-01T00:00:00" \
  xis_rmffile=suzaku/xis/cpf/ae_BI_ao4_20090901.rmf" \
  xis_contamifile="suzaku/xis/bcf/ae_xi1_contami_20061016.fits" \
  outfile="suzaku_xis_events.fits"
\end{lstlisting}
{\it XMM(EPIC):}
\begin{lstlisting}
runsixt \
  EventList="sixtxmm_events.fits" \
  PatternList="sixtxmm_pattern.fits" \
  Mission="XMM" Instrument="EPICPN" \
  Mode="FFTHIN" \
  XMLFile="xmm/epicpn/fullframe_thinfilter.xml" \
  Simput="phlist.fits" \
  Exposure=1.0e4E \
  RA=10.0 Dec=10.0 \
  MJDREF=50814.0
epicpn_events \
  PatternList="sixtxmm_pattern.fits" \
  EPICpnEventList="xmm_epic_events.fits"
\end{lstlisting}
{\it eROSITA:}
\begin{lstlisting}
erosim \
  prefix="erosita_" \
  PhotonList=events_pv.fits \
  RawData=events_allpv.fits \
  background=yes \
  XMLFile="srg/erosita_1.xml" \
  XMLFILE1="srg/erosita_1.xml" \
  XMLFILE2="srg/erosita_2.xml" \
  XMLFILE3="srg/erosita_3.xml" \
  XMLFILE4="srg/erosita_4.xml" \
  XMLFILE5="srg/erosita_5.xml" \
  XMLFILE6="srg/erosita_6.xml" \
  XMLFILE7="srg/erosita_7.xml" \
  Simput="phlist.fits" \
  Exposure=1.0e4 \
  SkipInvalids=yes \
  seed=-1 \
  clobber=yes \
  RA=10.0 Dec=10.0 \
  MJDREF=50814.0
\end{lstlisting}
{\it Athena(XIFU):}
\begin{lstlisting}
xifupipeline \
  prefix="athena_xifu_" \
  PixImpactList=impact.fits \
  XMLFile=athena/1469mm_xifu/xifu_baseline.xml \
  AdvXml=athena/1469mm_xifu/xifu_detector_hex_baseline.xml \
  Background=yes \
  RA=10.0 Dec=10.0 \
  Simput="phlist.fits" \
  Exposure=1.0e4 \
  UseRMF=T  \
  clobber=yes
\end{lstlisting}
{\it Athena(WFI):}
\begin{lstlisting}
athenawfisim \
  prefix=athena_wfi_ \
  XMLFile0="athena/1469mm_wfi_w_filter/depfet_b_1l_ff_chip0.xml"\
  XMLFile1="athena/1469mm_wfi_w_filter/depfet_b_1l_ff_chip1.xml"\
  XMLFile2="athena/1469mm_wfi_w_filter/depfet_b_1l_ff_chip2.xml"\
  XMLFile3="athena/1469mm_wfi_w_filter/depfet_b_1l_ff_chip3.xml"\
  Simput="phlist.fits" \
  Exposure=1.0e4 \
  Background=yes \
  Attitude="athena/attitude_wfi_ra10_dec10.fits" \
  RA=10.00 Dec=10.00 \
  chatter=0 \
  MJDREF=52000.0 \
  clobber=yes
\end{lstlisting}

{\it XraySurveyor(HDXI)}
\begin{lstlisting}
  pset simx mode=hl
  pset simx Exposure=1.0e4
  pset simx UseSimput=yes
  pset simx MissionName=XraySurveyor
  pset simx InstrumentName=HDXI
  pset simx ScaleBkgnd=0.0
  pset simx RandomSeed=24
  pset simx SimputFile=phlist.fits
  pset simx PointingRA=10.0
  pset simx PointingDec=10.0
  pset simx OutputFileName=surveyor_hdxi_events
  simx
\end{lstlisting}

{\it XraySurveyor(XCAL)}
\begin{lstlisting}
  pset simx mode=hl
  pset simx Exposure=1.0e4
  pset simx UseSimput=yes
  pset simx MissionName=XraySurveyor
  pset simx InstrumentName=XCAL
  pset simx ScaleBkgnd=0.0
  pset simx RandomSeed=24
  pset simx SimputFile=phlist.fits
  pset simx PointingRA=10.0
  pset simx PointingDec=10.0
  pset simx OutputFileName=surveyor_xcal_events
  simx
\end{lstlisting}

{\it Hitomi(SXS)}
\begin{lstlisting}
  pset simx mode=hl
  pset simx Exposure=1.0e4
  pset simx UseSimput=yes
  pset simx MissionName=Hitomi
  pset simx InstrumentName=SXS
  pset simx ScaleBkgnd=0.0
  pset simx RandomSeed=24
  pset simx SimputFile=phlist.fits
  pset simx PointingRA=10.0
  pset simx PointingDec=10.0
  pset simx OutputFileName=hitomi_sxs_events
  simx
\end{lstlisting}

{\it Hitomi(SXI)}
\begin{lstlisting}
  pset simx mode=hl
  pset simx Exposure=1.0e4
  pset simx UseSimput=yes
  pset simx MissionName=Hitomi
  pset simx InstrumentName=SXI
  pset simx ScaleBkgnd=0.0
  pset simx RandomSeed=24
  pset simx SimputFile=phlist.fits
  pset simx PointingRA=10.0
  pset simx PointingDec=10.0
  pset simx OutputFileName=hitomi_sxi_events
  simx
\end{lstlisting}

{\it Hitomi(HXI)}
\begin{lstlisting}
  pset simx mode=hl
  pset simx Exposure=1.0e4
  pset simx UseSimput=yes
  pset simx MissionName=Hitomi
  pset simx InstrumentName=HXI
  pset simx ScaleBkgnd=0.0
  pset simx RandomSeed=24
  pset simx SimputFile=phlist.fits
  pset simx PointingRA=10.0
  pset simx PointingDec=10.0
  pset simx OutputFileName=hitomi_hxi_events
  simx
\end{lstlisting}

{\it Chandra(ACIS-S)}
\begin{lstlisting}
  marx S-SIMPUT-Source="phlist.fits" \
  ExposureTime=1.0e4 TStart=2012.5 \
  GratingType="NONE" DetectorType="ACIS-S" \
  DitherModel="INTERNAL" RA_Nom=10 Dec_Nom=10 Roll_Nom=50 \
  SourceRA=10 SourceDEC=10 \
  Verbose=yes mode=h OutputDir=point
  marx2fits point chandra_acis-s_evt.fits
\end{lstlisting}

{\it Chandra(ACIS-I)}
\begin{lstlisting}
  marx S-SIMPUT-Source="phlist.fits" \
  ExposureTime=1.0e4 TStart=2012.5 \
  GratingType="NONE" DetectorType="ACIS-I" \
  DitherModel="INTERNAL" RA_Nom=10 Dec_Nom=10 Roll_Nom=50 \
  SourceRA=10 SourceDEC=10 \
  Verbose=yes mode=h OutputDir=point
  marx2fits point chandra_acis-i_evt.fits
\end{lstlisting}

\end{document}